# On peculiarities of superconductivity in the single-layer FeSe/SrTiO$_3$ interface

Lev P. Gor'kov[1,2]

[1]*NHMFL, Florida State University, 1800 E. Paul Dirac Drive, Tallahassee, Florida, 32310, USA*
[2]*L.D. Landau Institute for Theoretical Physics of the RAS, Chernogolovka, 142432, Russia*

Observation of replica bands in the ARPES spectra of the single-layer FeSe on strontium titanate substrate revealed a phonon component among mechanisms behind high $T_C$ superconductivity there. We study the interaction of the in-layer FeSe electrons with the electric potential of the longitudinal (LO) modes at surface of bulk SrTiO$_3$. A two-dimensional system of charges at the FeSe/SrTiO$_3$ interface includes both the itinerant and immobile electrons. The latter significantly change the interface characteristics, increasing screening at the substrate surface and reducing thereby the strength of the electron-LO phonon interactions. In that follows the dielectric constant serves as a free parameter and is found from the ARPES data for the replicas. The two-dimensional Coulomb screening is accounted for in the random phase approximation. It is shown that the model is applicable within the whole range of the parameters typical for current experiments. The estimates allow concluding that the LO-phonon mediated pairing alone cannot account for the temperatures of the superconducting transition $T_C$ in the single-layer FeSe/SrTiO$_3$ reported in these experiments. This does not exclude that the LO-phonons mechanism can become more significant in the differently and better prepared single layer FeSe films. Available experiments are briefly discussed. Thus far no data exist on the dependence of $T_C$ on the concentration of electrons doped into the in-layer FeSe band.

*Introduction.* The recent discovery of superconductivity in the single layer of FeSe deposited on the strontium titanate substrate (STO) [1] with the temperature of transition up to 110K [2] is of great interest on both practical and theoretical grounds. On one hand, it opens new prospects in applications, including, in particular, the engineering of interfaces and films possessing such property as superconductivity. On the theory side, the whole manifold of the experimental data gives evidence that a phonon mechanism is undoubtedly again at work challenging thereby the forty-years-old consensus in the literature regarding stringent limitations on the temperature of the superconducting transition achievable with the later [3].

In that follows, we undertake attempts to reveal the basic factors that control both the normal and superconductivity properties of the single-layer FeSe/STO. Disclosure of replica bands in the angle-resolved photoemission spectroscopy (ARPES) data [4, 5] made self-evident the fact of coupling between electrons and a high frequency surface phonon mode. In general terms, the idea that surface modes may be responsible for pairing between the in-layer FeSe electrons was of course put forward in the literature. To be concrete, we focus on the interaction of electrons with the longitudinal (LO) surface polar modes on the charged STO substrate.

As a matter of fact, this problem cannot be addressed without making some suppositions regarding the structure of the interface and the mechanisms responsible for the electronic doping of the FeSe-layer. In current experimental literature [1, 2, 4-9] the preliminary procedures at sample preparation consist in annealing the substrate in a vacuum in order to produce the oxygen vacancies and to form, thereby, charged two-dimensional layer at the SrTiO$_3$ surface. The unit-cell thick FeSe layer is then deposited by molecular beam epitaxy (MBE). We gather that transferred into the conduction band of the single FeSe layer after the deposition is only a fraction of the charges. The rest stays on the surface of SrTiO$_3$ [5-7, 9 and 10]. While solution of the problem of electrons interacting with the electric fields inherent in LO polar modes on the surface of a *dielectric* is well known [11], in the present case it becomes complicated by changes in the dielectric properties of the surface caused by the immobile charge remaining on the substrate. The presentation below is an attempt to account for these peculiarities in the phenomenological model generalizing the standard approach [11].

Although from observation of the replica bands [4] one can infer that the interaction between electrons and the surface optical phonons is among the key features in the system, there remains the question whether the mechanism of the LO phonon-mediated pairing alone can lead to so high $T_c$ as observed in the single-layer FeSe/STO (1UCFeSe/STO) [2]. In the frameworks of the suggested approach we derive the expression for the intensity of the replica bands that allows analyzing data [4] quantitatively. Within the range of parameters revealed by this analysis the answer is negative. The conclusion agrees with that one in [4]. Nevertheless, when acting in concert with other mechanisms, the contribution from the LO surface phonons found capable to significantly enhance $T_c$ compared to its value in bulk FeSe. Note in passing that the particular mechanism of superconductivity in the latter is currently under debate (see, e.g. [12, 13]).

*The model.* Assume for the start that the interaction of electrons with a high frequency optical surface mode is the sole mechanism of superconductivity in this system. At that, the main unconventional feature from the theory viewpoint is the inverted ratio between the Fermi $E_F$ and the characteristic LO phoning frequency. In traditional metals the typical frequency $\omega_0$ of the phonons contributing to pairing has the same order of magnitude as the Debye temperature. The latter usually equals few hundred degrees and is some two orders of magnitude smaller $E_F \sim 1 eV$. In turn, the temperature of transition is also rather small, from $T_C \simeq 7.2 K$ in Lead (Pb) that is of the order of one tenth $\theta_D$, and few Kelvin degrees $T_C \sim 10^{-2} \theta_D$ typical for most metals of the main groups. In the case in hand $E_F \approx 60 meV$, $\omega_0 \approx 80 meV$ and $T_C \approx (50 \div 100) K$ [2, 4 and 9].

The ratio $T_C / \omega_0 \approx 1/8$ (with $T_C \approx 110 K$ [2]) suggests that interactions in the system responsible for pairing are not too weak. In metals the extension of the BCS weak coupling model to the case of arbitrary strong interactions is realized by the set of the well-known Eliashberg equations [14]. However, these equations are applicable only in the so called adiabatic limit, that is, at the condition that the Migdal parameter $r = \omega_0 / E_F$ is small $\omega_0 \ll E_F$ [15].

With the Migdal adiabatic provision severely violated in the new system, the discussions below would inevitably acquire a qualitative character. In a crude approximation, however, one might formally consider the one eight in the ratio $T_C / \omega_0 \approx 1/8$ as due to a small enough coupling parameter $\lambda$ in the BCS-like expression for the temperature of superconducting transition:

$$T_c = const \times W(2\gamma / \pi) \exp(-1/\lambda) . \quad (1)$$

We find below that the electron-optical phonon interactions mechanism cannot explain $T_C \approx 100 K$, so indeed the corresponding coupling parameter $\lambda$ is small. Accordingly, for the single-layer FeSe/STO we adopt the idealized weak coupling model of the two-dimensional parabolic band of electrons at the $M$-point of the Brillouin zone (BZ) [4,5 and 9]. For the simplicity's sake, we assume at first the *extreme* "anti-adiabatic" case $\omega_0 \gg E_F$.

*Interaction of two-dimensional electrons with longitudinal surface optical phonons.* Interaction between electrons and the electric potential generated by LO phonons in polar crystals is described by the Fröhlich Hamiltonian:

$$\vec{P} = F_C \vec{u} \quad (2)$$

where $\vec{P}$ and $\vec{u}$ are the polarization and lattice displacement, respectively. In particular, the well-known case is the interaction between the surface optical phonons and electrons on the surface of a *clean* dielectric [11]. The coefficient $F_C$ equals:

$$\bar{F}_{C,i} = \left[ 4\pi e^2 \frac{\hbar \omega_{SLO}^i}{2} \left( \frac{1}{\kappa_\infty + 1} - \frac{1}{\kappa_0 + 1} \right) \right]^{1/2}. \quad (3)$$

$\kappa_0$ and $\kappa_\infty$ are the static and the optical dielectric constants of the bulk. In (2) $\omega_{SLO}^i$ is the frequency of one of the SLO phonon modes. The matrix element for the scattering of two electrons on each other via the virtual exchange by the surface phonon is:

$$M_i(q) = -\frac{4\pi e^2}{q} \left( \frac{1}{\kappa_\infty + 1} - \frac{1}{\kappa_0 + 1} \right) \times D_{SOP}^i(q). \quad (4)$$

In Eq. (3) $D_{SOP}^i(q)$ is the phonons Green function in the thermodynamic technique [16]:

$$D_{SOP}^i(q)(\varepsilon_n - \varepsilon_m) = \frac{(\omega_{SLO}^i)^2}{(\omega_{SLO}^i)^2 + (\varepsilon_n - \varepsilon_m)^2}. \quad (5)$$

In (3, 4) $\vec{q} = \vec{p} - \vec{k}$ and $\varepsilon_n - \varepsilon_m$ are the momentum and the frequency by which two electrons exchange upon scattering. In bulk, the well-known relation between $\omega_{LO}$ the frequency of the LO and the frequency $\omega_{TO}$ of the soft transverse optical phonons is $\omega_{LO}/\omega_{TO} = \sqrt{\kappa_0/\kappa_\infty}$; according to [11], from here, for the frequency $\omega_{SLO}$ of the surface phonon follows $\omega_{SLO}/\omega_{TO} = \sqrt{\kappa_0 + 1/\kappa_\infty + 1}$.

As it was already mentioned in the Introduction, for the problem in hand inconsistencies between experimental results and their interpretation seem to come about from incomplete understanding of the doping mechanisms. Without entering into excessive details of a customary doping protocol, it is worth to enumerate the main steps. A carefully prepared $TiO_2$-terminated $SrTiO_3$ substrate is annealed in vacuum at a high temperature, thereby creating the oxygen vacancies at its surface and forming a charged surface layer hosting a two-dimensional electron system on the titanium $3d$ $t_{2g}$–levels [1, 2, 4-9]. It is worth of note that the layer of FeSe is deposited by the molecular beam epitaxy (MBE) *after*, i.e., on top of *annealed surface*. A fraction of electrons from the charged $SrTiO_3$ surface goes over into the single-layer FeSe conduction band; yet, another part undoubtedly remains embedded in a thin surface layer on the substrate. Experimentally, it is firmly established that the results do not depend on whether SrTiO$_3$ is insulating in bulk or the Nb-doped. From this also follows the corollary that the conducting FeSe layer and the interfaces make a whole (see e. g. [5-7]).

The experimental discovery that, consistently with the above considerations, finalizes the model is the disclosure of a threshold in concentration for the doped carriers to appear at the chemical potential [6-8]. Such threshold signifies the existence of the mobility edge; only at concentrations above the threshold the carriers start manifesting themselves in the itinerant conductivity and superconductivity. Whether the concentration for the onset of the threshold can be controlled by a specific annealing protocol remains unclear from [8], but the very fact of the such threshold existence is critically important. At concentrations exceeding the threshold the itinerant and immobile carriers coexist and it becomes necessary to keep in mind that all phenomena in the conducting 1UCFeSe/STO-interface take place on a reconstructed dielectric background. As the dielectric constant $\kappa_0$ of SrTiO$_3$ is very large ($\kappa_0 = 1000$ at $T = 100K$ [17]), the local states below the mobility edge possess large dipole moments that contribute significantly into the polarization.

Below we *hypothesize* that the interactions between two-dimensional electrons and the optical phonons in 1UCFeSe/STO have the same form as given by the expressions (2- 4) with the difference that the parameters $\kappa_0$ and $\kappa_\infty$ must be redefined to account for the inevitable change in the dielectric characteristics of the SrTiO$_3$ surface at annealing and depositing the FeSe-layer . Therefore, from now on in all expressions $\tilde{\kappa}_0$ and $\tilde{\kappa}_\infty$ are the model parameters that may depend on details of the specific experiment.

Return to the expression of the matrix element for scattering of two electrons by each other via the virtual exchange by the surface LO phonon. Together with the term corresponding to common scattering of two electrons via the direct electron-electron Coulomb interaction, the total matrix element (3) acquires the form:

$$\mathrm{M}_{tot}(\vec{p},\varepsilon_n \mid \vec{k},\varepsilon_m) = \frac{4\pi e^2}{(\tilde{\kappa}_\infty + 1)q} - \sum_i \frac{4\pi e^2}{q(\tilde{\kappa}_\infty + 1)} V_i^2 \times D_{SLO}^i(\varepsilon_n - \varepsilon_m). \quad (6)$$

Summation is on the number $N$ of the optical phonons in the system. We assume $N = 3$, because there are *three* infrared-active LO phonon modes at the $\Gamma$-point of bulk SrTiO$_3$ [18], each with a frequency $\omega_{LO}^i > T_C$ [19]. The factor $V_i^2$ accounts for the fact that in the multi-mode polar crystals coupling of the optical phonons with electrons generally differs from that in Eq. (4) and the coefficients $V_i^2$ have a more complicated form than in (3, 4). Still, in SrTiO$_3$ among all phonon modes, one LO mode reveals the giant gap between its frequency and frequencies of all the transverse optical (TO) phonons [17, 19]. Therefore, the contribution from this mode into (6) can be taken as before in the same form as in Eq. (4). In particular, it compensates the direct Coulomb repulsion in Eq. (6) at $|\varepsilon_n - \varepsilon_m| << \omega_{LO}$.

The rest of LO phonons in (6) contribute to the matrix element for the electron-electron scattering. We emphasize that the latter corresponds to the *attractive* interaction. Besides, as $\kappa_0$ of SrTiO$_3$ is very large ($\kappa_0 >> \kappa_\infty$ [18]), we retain in the denominators (6) only the terms with the "optical" $\tilde{\kappa}_\infty + 1$.

We imply these considerations to the interaction between the band electrons in the FeSe layer and the LO phonons at the 1UCFeSe/STO interface. With the simplifying assumption of the *extreme* "anti-adiabatic" case $\omega_0 >> E_F$ the term $(\varepsilon_n - \varepsilon_m)^2$ in the denominator of the $D_{SLO}^i(\varepsilon_n - \varepsilon_m)$ can be

omitted. The matrix element of the interaction between the two electrons in the FeSe conduction band is:

$$\mathrm{M}_{tot}(\vec{p},\varepsilon_n | \vec{k},\varepsilon_m) \equiv M(\vec{p}-\vec{k}) \approx -2\alpha^2 \frac{4\pi e^2}{|\vec{p}-\vec{k}|(\tilde{\kappa}_\infty+1)} < 0. \quad (7)$$

Here the notation $2\alpha^2$ is introduced for the sum of $V_i^2$ over all other LO phonon modes; $\alpha^2 < 1$ is one of parameters in the model; the second parameter is $\tilde{\kappa}_\infty$ -the optical dielectric constant renormalized by the presence of electrons embedded into the interface layer.

The Coulomb interaction is screened by the two-dimensional gas of the FeSe-electrons. Restricting ourselves by the so called random phase approximation (RPA), the denominator in (7) becomes:

$$|\vec{p}-\vec{k}| \Rightarrow |\vec{p}-\vec{k}| + 4e^2 m / \hbar(\tilde{\kappa}_\infty+1). \quad (8)$$

Instead (7), one has:

$$M_{scr}(\vec{p}-\vec{k}) \approx -2\alpha^2 \frac{4\pi e^2}{\tilde{\kappa}_\infty} \times \frac{1}{|\vec{p}-\vec{k}| + 4e^2 m / \hbar(\tilde{\kappa}_\infty+1)}. \quad (9)$$

Generally, the 2D-screening depends on many structural details of the conducting layer [20]. Use of the RPA-type expressions (8, 9) can be justified in the "dense plasma" limit, i.e., when the kinetic energy of carriers prevails over the Coulomb interaction. In that case the inverse Thomas-Fermi radius $k_{TF} \equiv r_{TF}^{-1}$ must be small compared with a characteristic momentum $p_F$. For the former from Eq. (8) one finds $k_{TF} = 4e^2 m / \hbar(\tilde{\kappa}_\infty+1)$. For the experiments of interest below holds the following inequality:

$$p_F(\hbar(\tilde{\kappa}_\infty+1)/e^2 m) >> 1. \quad (10)$$

*Weak-coupling expression for temperature of the superconducting transition.* Temperature of the superconducting transition is determined by the eigenvalue of the homogeneous equation for the gap parameter $\Delta(\vec{p})$ (see [16] and the brief derivation in Appendix). In the notations (9) it reads:

$$\Delta(\vec{p}) = -T \sum_m \int \frac{d\vec{k}}{(2\pi)^2} M_{scr}(p-k) \Pi(k) \Delta(\vec{k}). \quad (11)$$

The Cooper instability originates from the logarithmic divergence related to blocks of the two Green functions $\Pi(k) \equiv G(k)G(-k) = [\varepsilon_m^2 + \varsigma^2]^{-1}$ in (11). (From now $\varsigma = (\vec{k}^2 - p_F^2)/2m \approx v_F(k-p_F)$). As it was pointed out above, at $\omega_{SLO} >> E_F$, dependence on the energy variable in (5) can be omitted $\mathrm{M}_{scr}(\vec{p},\varepsilon_n | \vec{k},\varepsilon_m) \Rightarrow M_{scr}(\vec{p}-\vec{k})$. Performing summation in (11) and substituting the explicit expression (9) for the integral kernel $M_{sc}(\vec{p}-\vec{k})$ results in the integral equation:

$$\Delta(\vec{p}) = \frac{2\alpha^2 e^2 m}{\pi \tilde{\kappa}_\infty} \int_{-\pi}^{\pi} \int_0^\infty \frac{d\varphi d\varsigma}{|\vec{p}-\vec{k}| + 4e^2 m / \hbar(\tilde{\kappa}_\infty+1)} \times \frac{1}{\varsigma} th \frac{\varsigma}{2T} \Delta(\vec{k}). \quad (12)$$

($\varphi$ is the angle between two vectors $\vec{p}$ and $\vec{k}$).

Let the vector $\vec{p}$ in (12) be *on the Fermi surface*. With the notation $\bar{\Delta}$ for an average value of $\Delta(k)$ rewrite the right hand side (12) as:

$$\frac{2\alpha^2 e^2 m}{\pi \tilde{\kappa}_\infty} \int_{-\pi}^{\pi} \int_0^W \frac{d\varsigma}{\varsigma} th\frac{\varsigma}{2T} \times \frac{d\varphi}{\sqrt{2p_F^2(1-\cos\varphi)+2m\varsigma}+4e^2 m/\hbar(\tilde{\kappa}_\infty+1)} \Delta(k) \Rightarrow \lambda \ln\left(\frac{W}{T}\right)\bar{\Delta}. \quad (13)$$

The integral over the angle $\varphi$ in front of the logarithmic singularity in Eq. (13) at $\varsigma = 0$ defines the exponential factor in the weak coupling expression $T_c = const \times W(2\gamma/\pi)\exp(-1/\lambda)$ for temperature of the transition. In (13) $W$ is a characteristic energy scale. That is, if the interaction kernel decreases at large energies, $W$ is the order-of-magnitude cutoff parameter in the integration over $\varsigma$ on the left in Eq. (13). (Strictly speaking, $\Delta(k)$ also depends of on the energy variable $\varsigma_k$). In our example of the *extreme* "anti-adiabatic" case $\omega_0 \gg E_F$ it is self-evident that $W = const \times E_F$. (The *exact* value of $const \sim 1$ can be determined only by solving the integral equation (12) explicitly).

Defining the effective Bohr radius $\bar{a}_B = (\tilde{\kappa}_\infty+1)\hbar^2/e^2 m$, inequality (10) reads in the new notation:

$$p_F \bar{a}_B \gg 1. \quad (14)$$

Introducing the dimensionless $x = (p_F \bar{a}_B / \hbar)/2$ and $\lambda$ in Eq. (13) denoting as $\lambda \equiv \alpha^2 \lambda(x)$ one obtains:

$$\lambda(x) = \frac{2}{\pi} \int_0^{\pi/2} \frac{du}{x\sin u + 1}. \quad (15)$$

For the temperature of transition:

$$T_c(x) = const \times (p_F^2/2m)\exp[-1/\alpha^2\lambda(x)] \equiv const \times (2\hbar^2/m\bar{a}_B^2)x^2 \exp[-1/\alpha^2\lambda(x)]. \quad (16)$$

For the purpose of illustration, take $\alpha^2 = 1$. The two functions $\lambda(x)$ and $t(x) = x^2 \exp[-1/\lambda(x)]$ are plotted in Fig. 1a, b. Maximum in $t(x)$ is obviously due to competition of the two factors: at a given $\bar{a}_B$ temperature of the transition initially increases with the increase of the concentration of carriers; at the same time, screening tends to reduce the effective constant of interaction $\lambda(x)$.

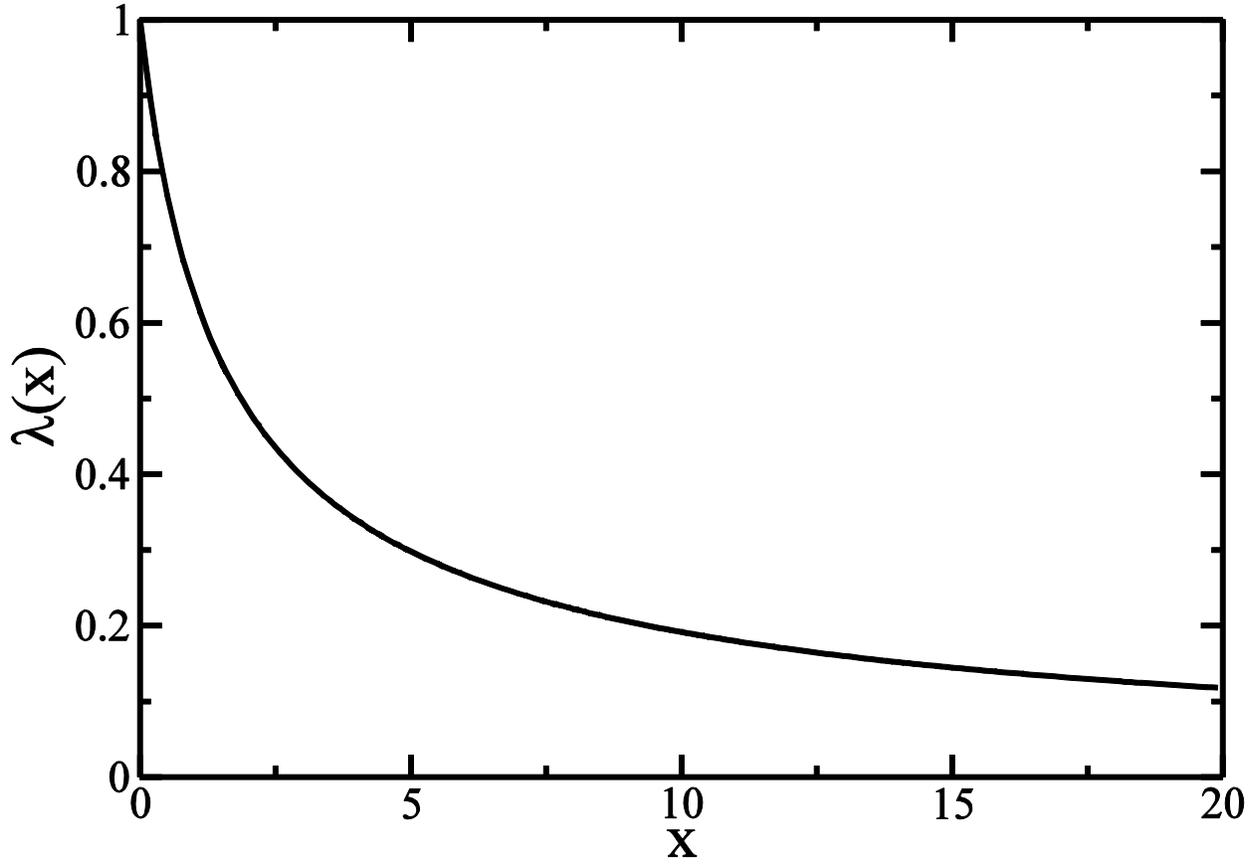

Fig.1a. The exponential factor in the weak coupling expression $T_c = const \times W \exp[-1/\lambda(x)]$ for the transition temperature, $\lambda(x)$ Eq. (15) as function of the dimensionless parameter $x = p_F \bar{a}_B / 2\hbar$; ($p_F$ -the Fermi momentum; $\bar{a}_B = (\bar{\kappa}_\infty + 1)\hbar^2 / e^2 m$ -the effective Bohr radius).

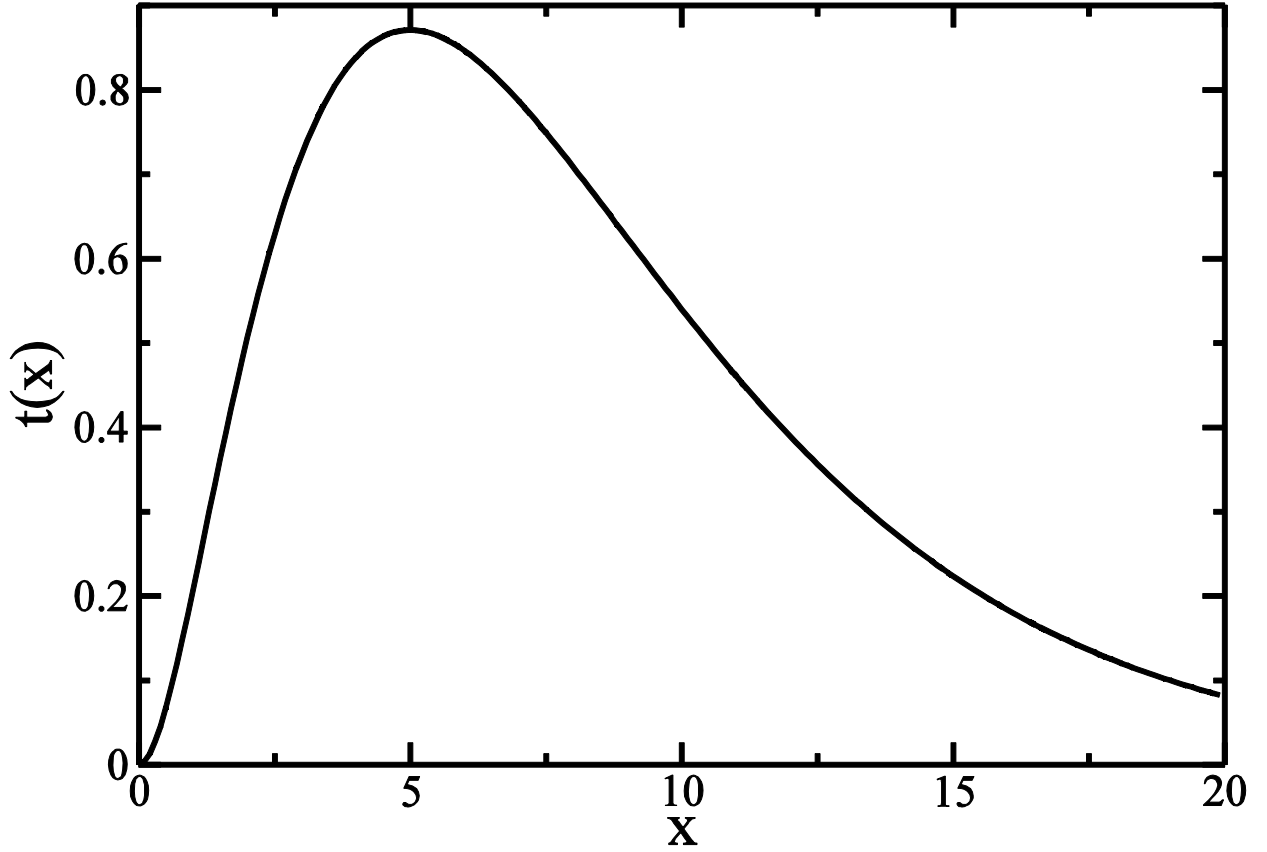

Fig.1b. The function $t(x) = x^2 \exp[-1/\lambda(x)]$. See the expression (16) for the temperature of superconducting transition $T_c(x) = const \times (p_F^2 / 2m) \exp[-1/\lambda(x)] \equiv const \times (2\hbar^2 / m\bar{a}_B^2) t(x)$.

*Replica bands.* The intensity of the ARPES spectra is proportional to the spectral function $A(\varepsilon, \vec{p})$ related to the imaginary part of the retarded Green function $G_R(\vec{k};\omega) = [\omega - \varepsilon(\vec{k}) + \mu - \Sigma(\vec{k};\omega)]^{-1}$ as:

$$A(\varepsilon, \vec{p}) = -\frac{1}{\pi} \frac{\mathrm{Im}\Sigma(\varepsilon, \vec{p})}{[\varepsilon - \varepsilon(\vec{k}) + \mu - \mathrm{Re}\Sigma(\varepsilon, \vec{p})]^2 + [\mathrm{Im}\Sigma(\varepsilon, \vec{p})]^2} . \quad (17)$$

In the general expression (17) for $A(\varepsilon, \vec{p})$ we confine ourselves to the intermediate states with one optical phonon. Correspondingly, for the self-energy in (17) we consider the diagram containing only one line of the phonon Green function. The analytical continuation of the expression $\Sigma(\varepsilon, \vec{p})$ on the thermodynamic axis:

$$\Sigma(\varepsilon, \vec{p}) = -T \sum_m \int \frac{d\vec{k}}{(2\pi)^2} M_{sc}(\vec{k}) G(\vec{p} - \vec{k}, \varepsilon - \varepsilon_m) D_{SLO}(\vec{k}, \varepsilon_m) \quad (18)$$

to the real frequency axis defines the retarded self-energy part $\Sigma_R(\varepsilon, \vec{p})$ as (see [16]):

$$\Sigma_R(\varepsilon,\vec{p}) = -\frac{1}{(2\pi)^3\pi}\int d^2\vec{k}M_{scr}(\vec{k})\int_{-\infty}^{+\infty}d\omega\int_{-\infty}^{+\infty}d\varepsilon_1 \frac{\mathrm{Im}G_R(\varepsilon_1,\vec{p}-\vec{k})\mathrm{Im}D_R(\omega,\vec{k})}{\omega+\varepsilon-\varepsilon_1-i\delta}\times[\tanh\frac{\varepsilon_1}{2T}+\coth\frac{\omega}{2T}] \quad (19)$$

Making use of the expressions for $\mathrm{Im}D(\omega,\vec{k}) = -\pi(\omega_{SLO}/2)\{\delta(\omega-\omega_{SLO})-\delta(\omega+\omega_{SLO})\}$ and for $\mathrm{Im}G_R(\varepsilon,\vec{p}) = -\pi\delta(\varepsilon-\varepsilon(\vec{p})+\mu)$, one obtains:

$$\Sigma_R(\varepsilon,\vec{p}) = \frac{\omega_{SLO}}{2^4\pi^2}\int d^2\vec{k}M_{scr}(\vec{k})\frac{1}{\omega_{SLO}-\varepsilon+\varepsilon(\vec{p}-\vec{k})-\mu+i\delta}\times[\tanh\frac{\mu-\varepsilon(\vec{p}-\vec{k})}{2T}+\coth\frac{\omega_{SLO}}{2T}]. \quad (20)$$

At low temperatures $T \ll E_F, \omega_{SLO}$ Eq. (20) simplifies to:

$$\mathrm{Im}\Sigma_R(\varepsilon,\vec{p}) = -(\omega_{SLO}/2^3\pi)\int d^2\vec{k}M_{scr}(\vec{k})\delta(\omega_{SLO}-\varepsilon+\varepsilon(\vec{p}-\vec{k})-\mu). \quad (21)$$

In practice, to improve resolution, the weak-intensity bands are usually analyzed by taking the second derivatives of the ARPES spectra. Following [4], consider the second derivative $(\partial^2/\partial\varepsilon^2)[\mathrm{Im}\Sigma_R(\varepsilon,\vec{p})]$ in Eq.(20). Rewriting $(\partial/\partial\varepsilon)\delta(\omega_{SLO}-\varepsilon+\varepsilon(\vec{p}-\vec{k})-\mu)$ as $-(1/v_x(\vec{p}-\vec{k}))(\partial/\partial k_x)\delta(\omega_{SLO}-\varepsilon+\varepsilon(\vec{p}-\vec{k})-\mu)$, repeating: $(\partial/\partial\varepsilon) \to -[1/v_y(\vec{p}-\vec{k})](\partial/\partial k_y)$, and leaving after the integration by parts only the most singular terms in $M_{scr}(\vec{k})$ Eq. (9), we find:

$$\frac{\partial^2}{\partial\varepsilon^2}\mathrm{Im}\Sigma_R(\varepsilon,\vec{p}) = -\frac{4\pi\omega_{SLO}e^2}{\tilde{\kappa}_\infty}\int\frac{kdk}{v^2(\vec{p}-\vec{k})}\times\frac{\delta(\varepsilon+\mu-\omega_{SLO}-\varepsilon(\vec{p}-\vec{k}))}{\{k+4e^2m/\hbar(\bar{\kappa}_\infty+1)\}^3}. \quad (22)$$

Assuming the integral converging at $4e^2m/\hbar(\bar{\kappa}_\infty+1) = 4\hbar/\bar{a}_B \ll p_F$ one finally obtains:

$$\frac{\partial^2}{\partial\varepsilon^2}\mathrm{Im}\Sigma_R(\varepsilon,\vec{p}) = -\frac{\pi\omega_{SLO}}{2v^2(\vec{p})m}\delta(\varepsilon+\mu-\omega_{SLO}-\varepsilon(\vec{p})). \quad (23)$$

*Discussion.* The experimental facts available thus far [1, 2, 5-8] are not specific to admit a detailed comparison with the theoretical results above, the more so, as no *quantitative* data on dependence of $T_C$ on the carriers concentration at the single-layer FeSe/STO were established with a degree of confidence (see e.g. [6-8]; the field effect has been observed in [6]).

The possibility for a more quantitative discussion presents itself in connection to the question regarding the origin of replicas [4].

There are two independent physical characteristics beside $\alpha^2$ that enter into the theoretical expressions (7, 9, 12 and 13): $p_F$ and the effective Bohr radius $\bar{a}_B = e^2m/\hbar^2(\bar{\kappa}_\infty+1)$. The values of $p_F$ are available directly from the ARPES spectra and define the surface density of electrons $n_s = (p_F^2/2\pi\hbar^2)$. The effective Bohr radius (and, hence, $\bar{\kappa}_\infty$) will be now evaluated indirectly making use of the ARPES data [4].

It was stressed in [4] that to account for the so accurate one-to-one correspondence between the dispersion of the electron energy bands and that of the replicas, especially at the $M$-point of the BZ, the electron-phonon interaction must be peaked at the small momentum transfer $|\vec{q}|$. Compare now $q_0 \approx 0.1 \times 10^8 cm^{-1}$, the experimental error bar for the replica-widths [4], with the value of cutoff $4e^2 m / \hbar(\bar{\kappa}_\infty + 1) = 4\bar{a}_B^{-1} \hbar$ in the denominator Eq. (22). The comparison gives $4\bar{a}_b^{-1} \simeq q_o$, or $\bar{a}_B \simeq 40 \times 10^{-8} cm$. With the band mass $m \approx 2m_e$ from [4] one finds $(\bar{\kappa}_\infty + 1) \simeq 160$ (in the *dielectric* SrTiO$_3$ $\kappa_\infty \simeq 5.2$ [18]).

At substitution of $p_F / \hbar \approx 0.3 \times 10^8 cm^{-1}$, the dimensionless parameter is $x = (p_F \bar{a}_B / \hbar) / 2 = 6$. The inequality (14) is fulfilled $p_F \bar{a}_B / \hbar = 12 \gg 1$. By that the applicability of the approach to the analysis above is justified.

From here one now obtains $\lambda(6) \approx 0.25$ (see Fig.1a). The substitution of this value into Eq. (15) leads to $T_c(x) \approx const \times E_F \times 0.02 = const \times 12K$ ( $E_F \approx 60 meV$ [4, 9]). Thereby, assuming $const \sim 1$, the interaction of electrons with LO optical phonons *alone* cannot account for $T_c = 58 \pm 7K$ reported in [4].

*If* there are *two* mechanisms contributing to the Cooper pairing, the gap equations (12, 13) must be rewritten:

$$\Delta(\vec{p}) = \int_0^{\tilde{E}} K(p,k) \frac{d\varsigma_k}{\varsigma_k} th \frac{\varsigma_k}{2T} \Delta_1(\vec{k}) + \frac{2\alpha^2 e^2 m}{\pi \tilde{\kappa}_\infty} \int_{-\pi}^{\pi} \int_0^{\infty} \frac{d\varphi d\varsigma}{|\vec{p} - \vec{k}| + 4e^2 m / \hbar \bar{\kappa}_\infty} \times \frac{1}{\varsigma} th \frac{\varsigma}{2T} \Delta(\vec{k}) \ . \quad (24)$$

In (24) $K(p,k)$ is now the kernel related to that specific pairing mechanism that, hypothetically, supports superconductivity in bulk FeSe. One view popular in the literature is that superconducting pairing in bulk FeSe is mediated by antiferromagnetic fluctuations (see, for instance, [13]). In case of such mechanism the characteristic cutoff energy in the first integral by the order of magnitude should be the same $\tilde{E} \sim E_F$. Assuming the weak-coupling expression $T_{C0} \approx E_F \exp(-1/\nu)$ in bulk FeSe, with $T_{C0} \simeq 8K$ and $E_F \approx 650K$ one finds from here $\nu = [\ln(E_F / T_{C0})]^{-1} \approx 0.23$. Adding $\lambda(x) \approx 0.25$ and $\nu \approx 0.23$ gives the total $\lambda_{tot} \approx 0.48$ in the exponent; at substitution into $T_C \approx E_F \exp(-1/\lambda_{tot})$ one finds for $T_C$ a reasonable estimate $T_C(x) \approx const \times 88K$.

If, instead being of the magnetic origin, $T_{C0} \simeq 8K$ in bulk FeSe were due to a commonplace phonon pairing with the Debye temperature $\theta_D \approx 200K$, for $\nu$ it would follow $\nu = [\ln(\theta_D / T_{C0})]^{-1} \approx 0.31$. As $\theta_D \ll E_F$, in this case Eq. (24) must be solved separately for $\Delta(\vec{k})$ in the two energy intervals $[0, \theta_D]$ and $[\theta_D, E_F]$. Simple calculations lead to the renormalized $\lambda_{ren}(x) = \lambda(x)[1 - \lambda(x) \ln(E_F / \theta_D)]^{-1} \approx 0.36$ and to $\lambda_{tot} \approx 0.67$; then one finds $T_C \approx const \times \theta_D \exp(-1/\lambda_{tot}) = const \times 45K$.

These estimates, although crude for the ultimate conclusions, not contradict the possibility that the record $T_C \simeq 109K$ [2] may be explained as due to the enhancement of same bulk $T_{C0} \simeq 8K$. On the theory side, it would be enough, as an example, to assume $x = (p_F \bar{a}_B / 2\hbar) \simeq 4$ and $\lambda(4) \approx 0.32$ (see in Fig. 1a). Note in passing that value $p_F \bar{a}_B / \hbar \simeq 8$ would satisfy the inequality (14) as well.

With two independent parameters $p_F$ and $\bar{a}_B$ there is only one dimensionless parameter in the theory $x = p_F \bar{a}_B / 2\hbar$. Post factum, from the above discussion one concludes that in the main part of the

$(p_F, \bar{a}_B)$-phase diagram in Fig.1a, b the use of RPA in Eqs. (8, 9) is warranted by the inequality (14). (Thus, a maximum of the function $t(x) = x^2 \exp[-1/\lambda(x)]$ in Fig.1b is at $x \simeq 5$ ($p_F \bar{a}_B / \hbar \simeq 10$)). At the level of current experiments [4] it may be possible to test Eq. (23). Namely, the second derivative of the replica band intensity (23) does not depend on $x$ while $T_c(x)$ decreases with the increase of $x$ (recall that ARPES can directly measure $p_F$).

The second parameter $(\bar{\kappa}_\infty + 1)$, intuitively, seem to be related to the particulars of the sample preparation procedure. At the fixed $x = p_F \bar{a}_B / 2\hbar \approx 6$ [4], reduction $\bar{a}_B$, say, by the factor three $(\bar{a}_B \to \bar{a}_B / 3)$ leads to $T_c = const \times 108 K$ in Eq. (16). For that density $n_s \approx 1.4 \times 10^{14} cm^{-3}$ in [4] must be increased to $n_s \approx 1.3 \times 10^{15} cm^{-3}$. This formal example is, however, an illustration that by establishing better control on mechanisms of doping one may manipulate the superconducting properties of the single-layer FeSe/STO.

*In summary,* we point out that with the sample preparations methods accepted in the current experimental literature the two-dimensional system of charges at the FeSe/SrTiO$_3$ interface inevitably includes both the itinerant and immobile electrons. Electrons trapped below the mobility edge are responsible for the change of the dielectric constant at the substrate surface. The Cooper pairing matrix elements in the single-layer FeSe/STO are calculated in the model of band electrons interacting with the electric potential of a longitudinal (LO) phonon mode at the SrTiO$_3$ surface. The dielectric constant at the surface is the free parameter of the model.

It is shown that the theoretical results are applicable in the whole range of the typical experimental parameters. In particular, screening of the Coulomb interaction can be accounted in the random phase approximation. The estimates for the temperature of transition lead to the conclusion that the LO-phonon mediated pairing alone cannot account for superconductivity at such temperatures as reported, for instance, in [4]. The conclusion, however, may not be the ultimate one, as the theoretical expressions in general, not contradict the possibility that with better control of the mechanisms of doping one can enhance further the superconducting properties of 1uCFeSe/STO.

**ACKNOWLEDGMENTS**

The author thanks T. Siegrist and C. Beekman for many helpful discussions and for the clarification of a number of significant experimental details. I am grateful to H. J. Mard for creating the graphic material. The work is supported by the National High Magnetic Field Laboratory through NSF Grant No. DMR-1157490, the State of Florida and the U.S. Department of Energy.

**Appendix**

The onset of superconductivity at the temperature of transition $T_C$ manifests itself in the occurrence of the pole in the scattering amplitude of two electrons with zero total momentum and frequency [23]. In the notation $\Gamma(p, -p | p', -p') \equiv \Gamma(p | p')$ the amplitude is the sum of all diagrams in the Cooper channel:

$$\Gamma(p,|p') = M(p-q) - \frac{T}{(2\pi)^2}\sum_{n'}\int d\vec{k} M(p-k)G(k)G(-k)\Gamma(k,|p') \ . \quad (4)$$

The temperature of the transition is determined via the eigenvalue of the following homogeneous equations. Substitution $\Gamma(p,|p') \to \psi(p)$ leads to the integral equation for a function $\psi(p)$:

$$\psi(p) = -T\sum_m \int \frac{d\vec{k}}{(2\pi)^2} M(p-k)\Pi(k)\psi(k) \ . \quad (5)$$

## References

[1] Q. -Y. Wang, Z. Li, W. -H. Zhang, Z. -C. Zhang, J.-S. Zhang, W. Li, H. Ding, Y. -B. Ou, P. Deng, K. Chang, J. Wen, C. -L. Song, K. He, J. -F. Jia, S. -H. Ji, Y. -Y. Wang, L. -L. Wang, X. Chen, X. -C. Ma, and Q. -K. Xue, Chin. Phys. Lett. **29**, 037402 (2012)
[2] J. -F. Ge, Z. -L. Liu, C. Liu, C. -L. Gao, D. Qian, Q. -K. Xue, Y. Liu, and J. -F. Jia, Nat. Mater. **14**, 285(2015)
[3] P. B. Allen and R. C. Dynes, Phys. Rev. B**12**, 905 (1975)
[4] J. J. Lee, F. T. Schmitt, R. G. Moore, S. Johnston, Y. -T. Cui, W. Li, M. , Z. K. Liu, M. Hashimoto, Y. Zhang, D. H. Lu, T. P. Devereaux, D. -H. Lee, and Z. -X. Shen, Nature, **515**, 245 (2014)
[5] R. Peng, H.C. Xu, S.Y. Tan, H.Y. Cao, M. Xia X.P. Shen, Z.C. Huang, C.H.P. Wen, Q. Song,T. Zhang, B.P. Xie, X.G. Gong, and D.L. Feng, Nat. Commun.**5**, 5044 (2014)
[6] W. Zhang, Z. Li, F. Li, H. Zhang, J. Peng, C. Tang, Q. Wang, K. He, X. Chen, L. Wang, X. Ma, and Q. -K. Xue, Phys. Rev. B **89**, 060506(R) (2014)
[7] X. Liu, D. Liu, W. Zhang, J. He, L. Zhao, S. He, D. Mou, F. Li, C. Tang, Z. Li, L. Wang, Y. Peng, Y. Liu, C. Chen, L.Yu, G.Liu, X. Dong, J. Zhang, C. Chen, Z. Xu, X. Chen, X. Ma, Q. Xue, and X.J. Zhou, , Nature Comm. **5**, 5047, 1(2014)
[8] J. He, X. Liu, W. Zhang, L. Zhao, D. Liu, S. He, D. Mou, F. Li, C. Tang, Z. Li, L. Wang, Y. Peng, Y. Liu, C. Chen, L. Yu, G. Liu, X. Dong, J. Zhang, C. Chen, Z.Xu, X. Chen, X. Ma, Q. Xue, and X. J. Zhou, PNAS **111**, 18501(2014)
[9] D. Liu, W. Zhang, D. Mou , J. He, Y.-B. Ou, Q.-Y.Wang, Z. Li, L. Wang, L. Zhao, S. He, Y. Peng, X. Liu, C. Chen, L. Yu, G. Liu, X. Dong, J. Zhang, C. Chen, Z. Xu, J. Hu, X. Chen, X. Ma, Q. Xue, and X.J. Zhou, Nat. Commun.**3**, 931 (2012)
[10] F. Zheng, Z. Wang, W. Kang, and P. Zhang, Sci. Reports, **3**, 02213 (2013)
[11] S. Q. Wang and G. D. Mahan, Phys. Rev. B **6**, 4517 (1972)
[12] Y. -Y. Xiang, F. Wang, D. Wang, Q.-H. Wang, and D.-H. Lee, *High-temperature superconductivity at the FeSe/SrTiO3 interface*, Phys. Rev. B **86**, 134508 (2012)
[13] D. -H. Lee, arXiv: 1508.02461v1 (2015)
[14] G.M. Eliashberg, Sov. Phys. JETP **11**, 696 (1960)
[15] A. B. Migdal, Sov. Phys. JETP **7**, 996 (1958)
[16] A. A. Abrikosov, L. P. Gor'kov, and I. E. Dzyaloshinskii, *Methods of Quantum Field Theory in Statistical Physics,* Prentice-Hall, Inc., Englewood Cliffs, New Jersey, 1963
[17] K. A. Muller and H. Bukard, Phys. Rev. B **19**, 3593 (1979)
[18] W. Zhong, R.D. King-Smith, and D. Vanderbilt, Phys. Rev. Lett.**72**, 3618 (1994)
[19] N. Choudhury, E. J. Walter, A. I. Kolesnikov, and C.-K. Loong, Phys. Rev. B 77,134111 (2008)
[20] F. Stern, Phys. Rev. Lett. **18**, 546 (1967)